\documentclass{article}

\usepackage[english]{babel}
\usepackage{amsmath, amssymb}
\usepackage{graphicx}

\newcommand{\re}{\mathrm e}
\newcommand{\ri}{\mathrm i}

\begin{document}

\title{Exact sampling of self-avoiding paths\\ via discrete Schramm-Loewner evolution}

\author{
  {Marco Gherardi} \\[0.2cm]
  {\it INFN --- Sezione di Milano I}\\
  {\it Universit\`a degli Studi di Milano} \\
  {\it Via Celoria 16, I-20133 Milano, Italy} \\
  {e-mail: {\tt Marco.Gherardi@mi.infn.it}}
}

\maketitle

\begin{abstract}
We present an algorithm, based on the iteration of conformal maps, that produces independent samples of self-avoiding paths in the plane.
It is a discrete process approximating radial Schramm-Loewner evolution growing to infinity.
We focus on the problem of reproducing the parametrization corresponding to that of lattice models, namely self-avoiding walks on the lattice, and we propose a strategy that gives rise to discrete paths where consecutive points lie an approximately constant distance apart from each other.
This new method allows us to tackle two non-trivial features of self-avoiding walks that critically depend on the parametrization: the asphericity of a portion of chain and the correction-to-scaling exponent.
\end{abstract}

\section{Introduction}

The self-avoiding walk (SAW) is the prototypical lattice model for polymer behavior \cite{DesCloizeauxJannink}.
It is defined as the uniform distribution over the set of all fixed-length nearest-neighbor walks on some lattice, such that no site
is visited more than once.
Self-avoidance reflects what are known as \emph{excluded volume} effects in polymer science \cite{Schaefer}.
The universal aspects of the SAW have been the subject of study for decades in the physical and mathematical literature.
While some of its properties are well understood and rigorously established (a thorough account can be found in \cite{MadrasSlade}),
it still poses some difficult (and some seemingly impossible) problems.
Mathematical and theoretical advances in the study of the SAW have always been paralleled by constant efforts for
devising new algorithms and numerical strategies \cite{GuttmannConway,Sokal:montecarlomethods,vanRensburg},
and since it is one of the simplest non-trivial models it can serve as a test ground for novel algorithms in polymer science.

A connection has been studied in two dimensions --- both numerically and analytically ---
between the critical SAW (in the limit where the number of steps goes to infinity)
and a continuum model, called Schramm-Loewner evolution (SLE).
SLE is a one-parameter family of stochastic processes in the complex plane producing 
random curves (traces) with conformal invariance ``built in''.
It has been conjectured that when the parameter is equal to $\frac{8}{3}$ the scaling limit of self-avoiding walks is obtained \cite{LSW:planarSAW} (this is the case we will be focusing on in this paper).
Later, numerical evidence has been given in favor of this correspondence, both in the half-plane \cite{Kennedy:lengthofanSLE} and in the whole-plane \cite{Gherardi} geometries.

What we propose here is an algorithm for sampling self-avoiding paths in the plane,
based on a discretized version of SLE.
Essentially, discrete paths are built by iterative composition of rotations together with 
one simple conformal map that takes a small circle and pulls a slit out of it.
Thanks to the way SLE works, it is possible to efficiently produce \emph{independent samples},
since the algorithm is based on simple Brownian motion,
which is very easy to sample.

Self-avoiding walks on the lattice have a natural parametrization,
which corresponds to counting the number of steps along the walk.
As long as one considers observables that depend only on the support
of the walk, the correspondence with SLE curves is well-understood.
But most of the quantities of interest in polymer physics do depend
on the labeling of points along the chain and can not be matched with
their SLE analogues, since SLE curves come with their own uncorrelated parametrization.
Actually, the
problem of finding a sensible definition of \emph{natural parametrization} for SLE curves is
still debated in the mathematical literature \cite{Lawler:reparam,LawlerSheffield:reparam}.
From a numerical point of view, one needs an affordable way of generating SLE samples
with the parametrization corresponding to the proper time of lattice models.
One such method was introduced and studied by Kennedy \cite{Kennedy:lengthofanSLE} and will be briefly reviewed in Section \ref{section:thechoiceofparametrization}.

We hereby introduce a new method, based on the observation that the SAW --- even when both the number of steps goes to infinity
and the lattice spacing goes to zero --- is such that the euclidean distance between two consecutive points on the chain is constant throughout the chain itself.
We require the same property for the SLE discrete trace $\left\{\gamma_n\right\}$, trying to attain an approximately constant \emph{step length}
\begin{equation}
\label{eq:constantsteplength}
\left|\gamma_n-\gamma_{n-1}\right| \approx \lambda .
\end{equation}
Discrete SLE chains are constructed by iterative composition of conformal maps, each one being responsible for producing a step 
$\gamma_{n-1}\rightarrow\gamma_n$.
The method we propose for keeping an approximately constant step length does so
by rescaling each step according to the Jacobian of the conformal map that acts on the corresponding segment.

We focus on whole-plane SLE, so that our algorithm explicitly concerns the SAW in the plane,
but the same reasoning and methods could be easily translated to other geometries, such as the chordal one.
For instance, minimal modification is needed in order to treat ensembles of self-avoiding walks
with fixed endpoints lying on the boundary of the domain.

Section \ref{section:discretewholeplanesle} is a brief introduction to the discrete process that we refer to as \emph{discrete whole-plane SLE}, which is the central object of interest lying at the heart of the algorithm;
Section \ref{section:thechoiceofparametrization} introduces the issues about the choice of parametrization;
Section \ref{section:thenumericalstrategy} describes the numerical strategy used to reproduce the natural parametrization of lattice models in the framework of discrete SLE; numerical results for the SAW are presented in Section \ref{section:results}: we measure the \emph{asphericity} of an inner portion of SAW --- which is a highly parametrization-dependent quantity --- and we discuss the first correction-to-scaling exponent.

\section{\label{section:discretewholeplanesle}Discrete whole-plane SLE}

We are not going to describe SLE here (the interested reader can find all the details in many excellent reviews, such as \cite{Lawler:book,Werner:stFlour,Cardy,KagerNienhuis,BauerBernard}).
The aim of this section is to present a discrete process approximating radial SLE growing to infinity, i.e. a measure on curves with one end-point in $z=1$ and the other at $\infty$, living on the complex plane minus the unit disc, $\mathbb C\setminus\mathbb D$.
A more in-depth presentation can be found in \cite{Gherardi}.
An analogous discrete process in the chordal geometry was introduced in \cite{Bauer:DSLE}, where its convergence to SLE was also studied.

Let us consider an ordered set of points
\begin{equation}
\gamma_n \in \mathbb C\setminus\mathbb D, \quad n=1,\ldots,N .
\end{equation}
We will call such a set \emph{trace} or \emph{chain}.
We are going to describe a stochastic process whose outcomes are such traces.
Let $\left\{\delta_j\right\}$ and $\left\{\Delta_j\right\}$ be two sequences of real numbers with $\Delta_j>0$.
Consider the maps
\begin{equation}
\label{eq:slitmap}
\phi^{\mathbb D}_j(z)=\frac{(z+1)^2-2\re^{-\Delta_j}z-(z+1)\left((z+1)^2-4\re^{-\Delta_j}z\right)^{1/2}}{2\re^{-\Delta_j}z} .
\end{equation}
These are conformal maps of $\mathbb D$ onto $\mathbb D\setminus \left[\phi^{\mathbb D}_j(1),1\right)$, whose action can
be described as growing a \emph{slit} $\left[\phi^{\mathbb D}_j(1),1\right)$ inside $\mathbb D$ along the real axis, of length
\begin{equation}
\label{eq:slitlength}
1-\phi^{\mathbb D}_j(1) = 1-\re^{\Delta_j}\left(2-\re^{-\Delta_j}-2\sqrt{1-\re^{-\Delta_j}}\right) .
\end{equation}
$\phi^{\mathbb D}_j$ is the inverse map of the solution at time $\Delta_j$ to the \emph{Loewner equation} in the disc
\begin{equation}
\partial_t f_t(z)= f_t(z)\frac{\exp\left(\ri a_t\right)+f_t(z)}{\exp\left(\ri a_t\right)-f_t(z)}, \quad f_0(z)=z ,
\end{equation}
in the special case where the \emph{driving function} is a constant $a_t\equiv 1$.
Notice that $\phi^{\mathbb D}_j$ is well-defined also on the boundary of $\mathbb D$.

\begin{figure}[t]
\centering
\includegraphics[scale=0.75]{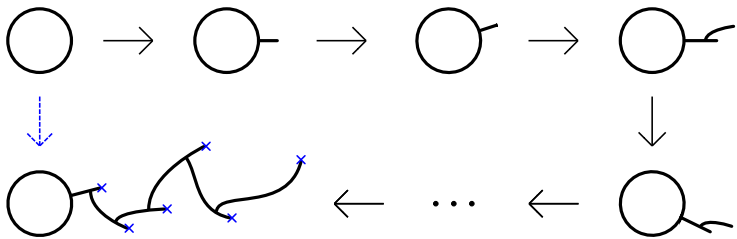}
\caption{(Color online) Composition of conformal maps giving rise to the trace.
The first (top-left) arrow corresponds to the slit map $\phi_n$, the second one (clockwise)
to the rotation $R_n$, the third one to $\phi_{n-1}$, the fourth one to $R_{n-1}$ and so on.
The last black arrow corresponds to the last rotation $R_1$.
The dashed blue arrow is the complete map $g_n$.
Blue crosses in the last picture identify the points $\left\{\gamma_n\right\}$ that constitute the trace.}
\label{figure:composition}
\end{figure}

By complex inversion we can then define a family of maps growing slits in the complement of $\mathbb D$ in $\mathbb C$:
\begin{equation}
\label{eq:wholeplanemap}
\phi_j(z)=\frac{1}{\phi^{\mathbb D}_j\left(1/z\right)} .
\end{equation}
The conformal maps $\phi_j$
send $\mathbb C\setminus\mathbb D$ onto $\mathbb C\setminus\mathbb D$ minus a slit on the real line,
whose length is controlled by $\Delta_j$\footnote{
See the discussion in Section \ref{section:thenumericalstrategy} for the relation between $\Delta_j$ and the length of the slit.}.
Let us construct a trace by intertwining such maps with rotations of the complex plane. 
We will consider the images of the point $z=1$ under the chain of maps obtained by alternately composing rotations and slit mappings, as follows:
\begin{equation}
\label{eq:tip}
\gamma_n=R_1\circ\phi_1\circ R_2\circ\phi_2\circ\cdots\circ R_n\circ\phi_n(1) ,
\end{equation}
where
\begin{equation}
R_j(z) = z\exp\left(\ri\delta_j\right)
\end{equation}
are rotations whose angles are the parameters $\delta_j$.
Let us call $g_n$ the $n$-th composed map, for use in the next sections:
\begin{equation}
\label{eq:nthmap}
g_n(z)=R_1\circ\phi_1\circ R_2\circ\phi_2\circ\cdots\circ R_n\circ\phi_n(z) ,
\end{equation}
so that $\gamma_n = g_n(1)$.
In words, we traverse the sequences $\left\{\Delta_j\right\}$ and $\left\{\delta_j\right\}$ backwards from $n$ to $1$
and for each $j$ we compose a slit mapping of parameter $\Delta_j$ with a rotation of angle $\delta_j$.
Refer to Figure \ref{figure:composition}.
At first a single slit is grown, and the point $z=1$ gets mapped onto the real axis, some distance away from the disc, namely at $\phi_n(1)$.
Then the universe is rotated and another slit is grown by application of $\phi_{n-1}$, so that the base of the previous slit --- which still lies on the unit circle after the rotation --- will be sent somewhere on the new slit\footnote{
We suppose for clarity that the rotations are small enough for this to be the case,
but even if the base of the slit remains on the circle the procedure
explained here produces nonetheless similar sets of points $\{\gamma_n\}$.}.
Notice that in general the shape of the old slit gets distorted because of the action of $\phi_{n-1}$.
The process goes on until the first map is reached; at that time, a chain of $n$ points has been produced.
Notice that, since the composition in (\ref{eq:tip}) goes backwards from $n$ to $1$, adding a step on the tip of the trace 
without changing the rest of the trace itself
means inserting $\phi_{n+1}$ at the rightmost place in (\ref{eq:tip}) and then recomputing the whole chain\footnote{
In the following we will mostly use 
letter $n$ to denote an index running from $1$ to $N$
(labeling the points on the chain)
and letter $j$ to denote an index running from $n$ to $1$
(labeling the incremental maps that build up the $n$-th composed map), but this is not a strict distinction, since they actually label the same sequences.
}.

The trace obtained of course depends on the two sequences $\left\{\delta_j\right\}$ and $\left\{\Delta_j\right\}$, so that a measure on the latter
induces a measure on the former.
We will draw $\delta_j$ and $\Delta_j$ in such a way that their relation be that of space versus time
for the one-dimensional rescaled Brownian motion.
The easiest way of doing so is to draw the $\delta_j$'s as Bernoulli variables in the set $\left\{\sqrt{\kappa\Delta_j},-\sqrt{\kappa\Delta_j}\right\}$ ---
where $\kappa$ is a positive real number --- and we shall do so in the following.
We still have the freedom to choose the ``time steps" $\Delta_j$.
A careful choice of the latter is what allows us to reproduce the parametrization of the lattice models.

The discrete process we have introduced here is expected to correspond to whole-plane SLE with parameter $\kappa$
in the appropriate limit (essentially, $N\to\infty$), and in particular to the SAW when $\kappa=\frac{8}{3}$.
Notice that the presence of the unit disc as a forbidden region is expected to become irrelevant if one looks at the trace sufficiently far away from the origin, so it is not surprising that the lattice counterpart is \emph{full-plane} SAW.

\section{\label{section:thechoiceofparametrization}The choice of parametrization}

The parameters $\left\{\Delta_j\right\}$ defined in the previous section are the time steps of the discretization.
They represent the time at which the Loewner evolution with constant driving function is to be evaluated in order to produce the
slit maps defined in (\ref{eq:slitmap}), which are the building blocks of the discrete evolution.
If one is interested in reproducing the actual SLE, where time is indeed a continuous variable, one will want to
have these parameters scale to zero.
The choice of \emph{how} they do so entails a choice of parametrization on the resulting object.
For instance, taking a constant $\Delta_j \equiv \Delta$ and then sending $\Delta$ to zero
yields SLE parametrized \emph{by capacity}, which means that the curves have linearly increasing capacity\footnote{
The (logarithmic) capacity is defined as the logarithm of the coefficient
of the term $z^{-1}$ in the expansion around $\infty$ of $g_n(z)$.}.
Different definitions can give rise to different parametrizations.
This is not an issue when one is interested only in parametrization-independent features of the curves, such as the fractal dimension,
the multifractal spectrum, the distance of the curve from some given point, or the probability of passing 
on the left or right of an obstacle, to name a few.
But it becomes crucial when one focuses on parametrization-dependent observables of the lattice models.
Such are for instance the distribution of a given point inside a SAW, the gyration tensor,
properties related to the detailed shape of the walks and
the universal quantities that describe the approach of an $N$-step chain to $N=\infty$,
such as the correction-to-scaling exponents (see Section \ref{section:results}).
Our goal is to find a choice of $\left\{\Delta_j\right\}$ capable of reproducing the natural parametrization of the lattice walk models.

The first thing one notices when producing discrete chains with the algorithm described in Section \ref{section:discretewholeplanesle}
is that the parametrization by capacity yields points that get further and further away from each other as the
process goes on.
The \emph{step size}
\begin{equation}
\label{eq:stepsize}
l_n = \left| \gamma_n-\gamma_{n-1} \right| 
\end{equation}
diverges when $n\to\infty$.
One can then choose to scale the time steps in such a way as to compensate for this.
It turns out (see \cite{Gherardi} for more details) that choosing
\begin{equation}
\label{eq:naivescaling}
\Delta_j \sim j^{-1} ,
\end{equation}
at least definitively in $j$, cancels out the drift in the distance between consecutive points.
Notice that this choice is \emph{non-random}, meaning that the reparametrization scheme does not depend
on the realization of the stochastic process.
Instead, the values of $\left\{\Delta_j\right\}$ are chosen \emph{before} the actual simulation takes place.
Unfortunately, this strategy does not give the correct parametrization 
(see also \cite{Kennedy:reparam}, where the same reasoning is applied to the half-plane case).
For instance, as far as the spatial distribution of the $k$-th point (for a given $k$) along the chain is concerned,
it gives exactly the same results as the parametrization by capacity.
What happens is that the scaling form (\ref{eq:naivescaling}) only ensures that the \emph{average}
distance $\left< l_n \right>$ between consecutive points be constant
(the average is over the realizations of the stochastic process).
But fluctuations around this average still retain all their correlations, since we are reparametrizing
in a naive, non-random fashion.

One solution to this problem was proposed and studied by Kennedy \cite{Kennedy:lengthofanSLE,Kennedy:reparam}, 
and is particularly adapted to the case when one is interested in the position of just a single point along the curve.
The idea is to grow the trace with its parametrization by capacity
and stop the growth when a fixed ``length'' has been reached.
A definition of length for a discretized fractal object can be introduced,
which turns out to be naturally dependent on a fixed length $\Lambda$, that is the scale
the fractal length (or \emph{variation}) is measured at.
Some care must be taken when using this method, because of the dependence of the results on the choice of the scale $\Lambda$.
One would like to send $\Lambda$ to zero, in order to measure the variation at a finer and finer mesh,
but at the same time $\Lambda$ can not become too small as compared to the step size of the discretized trace,
otherwise wilder and wilder rounding problems would completely spoil the computation.
Moreover, one wants to send the total length of the curve to infinity, which by scale invariance
amounts to shrinking the unit disc down to a point, so as to approach the truly whole-plane geometry.
One is therefore confronted with a tricky double limit, which increases the effort to be put into the analysis
of the numerical data, and can blur the estimation of the errors.

The method based on fractal variation is especially suited for producing a \emph{single} point on the chain at a given value of the parametrization.
The strategy we shall adopt here is different.
We aim at producing a discrete trace where the step sizes $l_n$ are strictly constant throughout the chain.
Stopping the discrete growth after a fixed number of steps will then be automatically equivalent to choosing the stopping time
when a fixed value of the fractal variation is reached.
The advantages of this strategy are manifold:
one obtains an essentially arbitrary number of points equally spaced in the natural parametrization
at the same cost as producing only the last one, and no computationally-delicate double limit is present, so that no additional
scaling analysis must be performed.
Moreover, no prior knowledge of the fractal dimension $d$ is needed.

\section{\label{section:thenumericalstrategy}The numerical strategy}

A close relative to the slit mapping in (\ref{eq:slitmap}) first appeared in the literature about \emph{diffusion-limited aggregation}, or DLA.
DLA --- introduced by Witten and Sander \cite{WittenSander:DLA} ---
is a kinetic model where finite-sized particles perform random walks (one at a time) from infinity until they stick irreversibly to a cluster, which grows from a seed placed at the origin.
Hastings and Levitov \cite{HastingsLevitov:DLA} took advantage of the conformal symmetry inherent to this model 
and proposed an algorithm which turns out to be similar to what we use for simulating Schramm-Loewner evolutions.
The algorithm works as follows.
The seed of the growth is the unit disc.
At each time-step, an angle $\theta_j$ is chosen with the uniform distribution in $[0,2\pi)$.
A conformal map $\phi_{A_j,\theta_j}$ is applied, that creates a bump of fixed area $A_j$ centered at $e^{\ri\theta_j}$.
Then, another $\theta$ is chosen, the maps are composed, and so forth.
Note that --- as is the case for Equation (\ref{eq:nthmap}) --- if $g_{n-1}$ is the map that grows the cluster
up to the $(n-1)$-th deposed grain, then the map that grows the cluster up to the $n$-th grain is
obtained by \emph{first} applying the incremental map $\phi_{A_n,\theta_n}$ and \emph{then} $g_{n-1}$, 
which is to say that the incremental maps are composed in the opposite order than usual.

This growth process satisfies an even stronger version of the domain Markov property\footnote{
It is essentially the usual Markov property, but ``up to'' a (time- and realization-dependent) conformal
transformation of the domain.}, 
which is one of the crucial characteristics of SLE \cite{Schramm},
since now the growth of the cluster at a specific time does not even depend on where it last grew, 
so not only does the future not depend on the past, but --- modulo a conformal transformation --- it does not depend on the present either.

An important technical aspect of this algorithm is that one wants to grow bumps of approximately equal size.
But peripheral bumps have undergone several conformal maps and have thus changed their shape and size to a great amount.
In general, by the time the whole cluster has been built, the $n$-th bump created (by $\phi_{A_n,\theta_n}$) has been subject to the action of $g_{n-1}$.
To compensate for this rescaling, one wants to create bumps of different sizes, depending on the whole history of maps they will be subject to in the remainder of the growth process.
As a first approximation, as long as the new ($n$-th) bump is sufficiently small, it is natural to try and correct 
only for the Jacobian factor $\left|g'_{n-1}\right|$ of the previous composed map,
calculated at the place where the new bump is to be created, because this is the rescaling factor that will affect the shape
of the bump at first order in its size.
The $n$-th bump size should then be
\begin{equation}
\label{eq:DLAfirstderivative}
A_n=\frac{A_0}{\left|g'_{n-1}(\re^{\ri\theta_n})\right|}.
\end{equation}
Since this strategy seems to give satisfactory results, it is very natural to try and apply 
it to the numerical reparametrization of SLE: rescaling the step sizes 
of the approximated SLE trace by the dilatation factor given by the Jacobian
(very similar ideas were also fruitfully exploited in \cite{Hastings:multifractal},
where multifractal spectra for Laplacian walks are computed).

The size $L_n$ of the $n$-th slit grown is a function of the time-like parameter $\Delta_n$ which controls the capacity of the incremental map at step $n$:
\begin{equation}
\label{eq:steplengthDelta}
\Delta_n = \log \frac{\left(2-\frac{L_n}{1+L_n}\right)^2}{4\left(1-\frac{L_n}{1+L_n}\right)},
\end{equation}
as can be seen by inverting (\ref{eq:slitlength}).
One wants to rescale $\Delta_n$ so that $L_n$ gets rescaled by a factor given by the Jacobian
\begin{equation}
\label{eq:Jacobian}
J_n = \left|g'_{n-1}\left(\re^{\ri\delta_n}\right)\right| ,
\end{equation}
in analogy with (\ref{eq:DLAfirstderivative}).

Unfortunately, there happens to be a great obstruction to this program, due to the fact that SLE satisfies ``only'' domain Markov property, instead of the complete independence of DLA steps that we discussed above.
If we look at Equation (\ref{eq:DLAfirstderivative}) we see that rescaling the step-sizes is possible because of the independence of $\theta_n$ (the space-like variable) from $A_n$ (the time-like variable).
This independence in DLA stems from the fact that the distribution of the $\theta_n$'s is flat on $[0,2\pi)$ and does not change, so that one can operatively choose every step $\theta_0,\theta_1,\ldots$ \emph{before} performing the composition of the corresponding maps.
In SLE, on the contrary, despite the fact that the steps satisfy the domain Markov property, the increments $\delta_n$ are drawn with a Bernoulli distribution from the set $\left\{\sqrt{\kappa\Delta_n},-\sqrt{\kappa\Delta_n}\right\}$, which does depend on time, since it depends explicitly on the time-like parameter $\Delta_n$; on the other hand the Jacobian needed to rescale $\Delta_n$ is to be evaluated at $\exp(\ri\delta_n)$.
Therefore, the problem is that we do not really know where to compute the Jacobian, until we have actually computed it!
This is ultimately related to the fact that SLE is driven by a non-trivial stochastic process, so that $\Delta_n$ and $\delta_n$ are intertwined.

\begin{figure}
\centering
\includegraphics[scale=0.8]{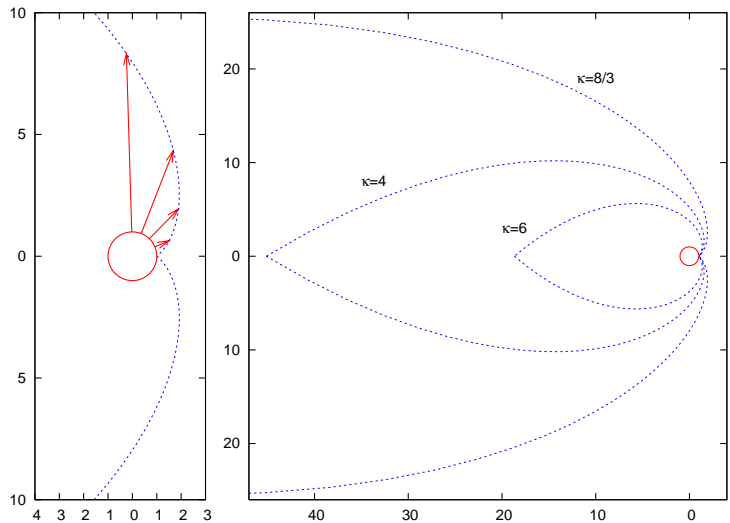}
\caption{(Color online)
The locus of the points $R_n\circ\phi_n(1)$ (dashed blue lines), 
as $\delta_n$ and $\Delta_n$ take on their allowed values.
The red circle is the unit disc;
the red arrows are examples of the possible slits grown.}
\label{figure:hearts}
\end{figure}
The reader can find a depiction of how the length of the slit $L_n$ and the angle $\delta_j$ are related in Figure \ref{figure:hearts}, which is a polar graph (for $-\pi<\delta_n<\pi$) for the position of the tip of the slit as a function of the angle  --- as can be found by inverting (\ref{eq:steplengthDelta}):
\begin{equation}
\label{eq:Ldelta}
L_n(\delta_n)=\frac{\exp\left(-\delta_n^2/\kappa\right)}{2 - \exp\left(-\delta_n^2/\kappa\right) - 2\sqrt{1-\exp\left(-\delta_n^2/\kappa\right)} } - 1 .
\end{equation}

Thus, the main problem with the foregoing approach is that $\delta_n$ and $\Delta_n$ depend on one another, so that one does not know where to compute the derivative.
One way to overcome this problem is the following.
Expand the derivative of $g_{n-1}$ (the map that grows the hull at step $n-1$) around its zero, which occurs at $z=1$, and evaluate it at the point $\exp(\ri\delta_n)$, which is the point where the $n$-th slit is going to be placed:
\begin{equation}
g'_{n-1}(\re^{\ri\delta_n})= g'_{n-1}(1) + \left(\re^{\ri\delta_n}-1\right) g''_{n-1}(1) + \ldots
\end{equation}
This expression is accurate when $\delta_n$ is small.
On the other hand, we also want to approximate the change in length
of the slit by the value of the derivative at the base,
silently assuming that it does not change much along the slit.
This approximation is justified by the fact that, by (\ref{eq:Ldelta}),
$L_n(\delta_n)$ is proportional to $\delta_n$ for $\delta_n$ small.

By expanding the exponential, taking the modulus, and remembering that $g'_n(1)=0$ for every $n$ ---
as can be seen by taking the derivative of (\ref{eq:nthmap}) with $\phi_n$ given by (\ref{eq:wholeplanemap}) and (\ref{eq:slitmap}) ---
one obtains the Jacobian (\ref{eq:Jacobian}) at order $\left|\delta_n\right|$
\begin{equation}
J_{n-1} \approx \left|\delta_n\right| \left|g''_{n-1}(1)\right| .
\end{equation}
We want to rescale the length $L_n$ of the $n$-th slit by $J_{n-1}$, so we rewrite 
the equation relating $\Delta_n$ and $L_n$ (\ref{eq:steplengthDelta}) by substituting
\begin{equation}
L_n = \frac{\lambda}{\left|\delta_n\right| \left|g''_{n-1}(1)\right|} ,
\end{equation}
where $\lambda$ is the desired step length (as in (\ref{eq:constantsteplength}), which represents our goal), 
and by making use of the Brownian relation
\begin{equation}
\left|\delta_n\right| = \sqrt{\kappa\Delta_n} .
\end{equation}
We obtain an equation which (if solved) gives the time-step $\Delta_n$ producing both the correct rescaling,
at first order in $\delta_n$, and the right relation with the space-step $\delta_n$:
\begin{equation}
\label{eq:transcendental}
\re^{\Delta_n} = \frac{1}{4} \left(2-\frac{\lambda}{\lambda+ \left|g''_{n-1}(1)\right|\sqrt{\kappa\Delta_n}} \right)^2
\left(1 - \frac{\lambda}{\lambda+ \left|g''_{n-1}(1)\right|\sqrt{\kappa\Delta_n}}\right)^{-1}.
\end{equation}
The actual sign of $\delta_n=\pm\sqrt{\kappa\Delta_n}$ is to be chosen at random, according to the Bernoulli nature of $\delta_n$.

Unfortunately, Equation (\ref{eq:transcendental}) is transcendental, and can not be solved explicitly.
A little thinking shows that for $\left|g''_{n-1}(1)\right|$ large one expects a small $\Delta_n$.
In fact, (\ref{eq:transcendental}) implies that the combination $\left|g''_{n-1}(1)\right|\sqrt{\kappa\Delta_n}$ be divergent when $\Delta_n\to 0$.
A crude approximation is then obtained by expanding the left hand side in powers of $\Delta_n$ around $0$, the right hand side in $\left|g''_{n-1}(1)\right|\sqrt{\kappa\Delta_n}$ around $\infty$ and matching the two behaviors at first order.
This is the best one can do, since higher orders would require solving algebraic equations of order greater than $4$ in $\sqrt{\Delta_n}$.
Among the solutions we choose the positive one:
\begin{equation}
\label{eq:Deltaj}
\Delta_n = \frac{\lambda}{2\sqrt{\kappa}\left|g''_{n-1}(1)\right|} .
\end{equation}
Numerical solution to (\ref{eq:transcendental}) shows that (\ref{eq:Deltaj}) is off by
$\sim 20\%$ when $\left|g''_{n-1}(1)\right|=10$ and by $\sim 5\%$ when $\left|g''_{n-1}(1)\right|=100$,
for $\lambda=1$ and $\kappa=\frac{8}{3}$.
The approximation works (see Section \ref{section:results}) because the typical values
of $\left|g''_{n-1}(1)\right|$ involved are large.

Of course, the foregoing method can be used only if one has an effective means of computing 
the main ingredient: $\left|g''_{n-1}(1)\right|$.
It turns out that there is such a way.
Straightforward calculations show (the details are in the Appendix) that
\begin{equation}
\label{eq:secondderivativeexpression}
\left|g''_n(1)\right|=\left|\phi''_n(1)\right| \prod_{j=0}^{n-2}\left|\phi'_{n-1-j}\left(\Gamma_j\right)\right| ,
\end{equation}
where $\Gamma_j$ is defined as
\begin{equation}
\label{eq:Gamma}
\Gamma_j = R_{n-j}\circ\phi_{n-j}\circ R_{n-j+1}\circ\phi_{n-j+1}\circ\cdots\circ R_n\circ\phi_n(1) .
\end{equation}
Equation (\ref{eq:secondderivativeexpression}) is a closed formula for the Hessian, in terms of $\left|\phi''_n(1)\right|$ ---
which only depends on $\Delta_n$, see (\ref{eq:phisecond}) ---
and the function $\phi'$ --- see (\ref{eq:phiprime1}) and (\ref{eq:phiprime2}).
Notice that the points $\Gamma_{j=0,\ldots,n-2}$ must already be computed by the routine that produces
the $n$-th point on the trace $\gamma_n=R_1\circ\phi_1\circ\cdots\circ\ R_n\circ\phi_n(1)$ ---
as was explained in Section \ref{section:discretewholeplanesle} ---
so that computing $\left|g''_n(1)\right|$ adds very little computational load.
At the $n$-th step of the algorithm --- i.e. when producing the $n$-th point along the chain --- 
one can compute the factor $\left|g''_n(1)\right|$,
which will be needed for producing the $(n+1)$-th point, simply by multiplying the constant $\left|\phi''_n(1)\right|$
together with all factors $\left|\phi'_{n-1-j}(\Gamma_j)\right|$
obtained at each composition that is performed to compute $\gamma_n$.
This amounts to performing an operation taking a time $O(1)$ for each composition,
which sums up to $O(n)$ for the $n$-th point --- that requires $n$ compositions ---
and finally to $O(N^2)$ for a complete $N$-step chain.

Let us schematically sum up how our algorithm for building an $N$-step chain works:

\begin{enumerate}
\item Set the constant $\lambda$ (we shall always fix $\lambda=1$)
\item Set $n\leftarrow 1$ and $\Delta_1\leftarrow 1$
\item \label{enum:outercyclestart} Compute $\delta_n$ as $\pm\sqrt{\kappa\Delta_n}$ with a random sign
\item Setup a temporary variable $D\leftarrow\left|\phi''_n(1)\right|$ calculated as in (\ref{eq:phisecond})
\item Set $j\leftarrow n$ and $z\leftarrow 1$
\item \label{enum:innercyclestart}Cycle on $j$ for computing $\gamma_n$:
	\begin{enumerate}
	\item Apply $j$-th incremental map and rotation $z\leftarrow R_j\circ\phi_j(z)$
	\item If $j>1$ multiply $D$ by $\left|\phi'_{j-1}(z)\right|$ calculated as in (\ref{eq:phiprime1}) and (\ref{eq:phiprime2})
	\item If $j>1$ decrease $j$ by $1$ and repeat step \ref{enum:innercyclestart}
	\end{enumerate}
\item Set $\gamma_n\leftarrow z$
\item Compute $\Delta_{n+1}$ as in (\ref{eq:Deltaj}) with $\left|g''_n(1)\right|$ given by $D$
\item If $n<N$ increase $n$ by 1 and go back to step \ref{enum:outercyclestart}
\end{enumerate}

\begin{figure}[t]
\centering
\includegraphics[scale=0.8]{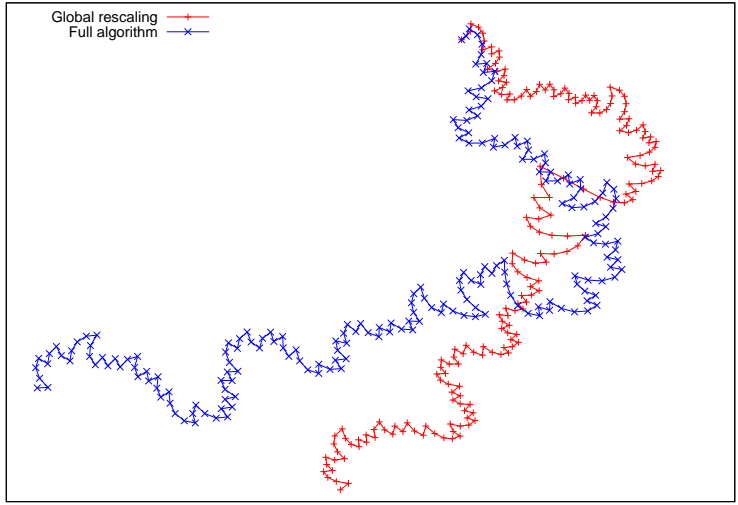}
\caption{(Color online) An example of a trace obtained by the full algorithm 
described in Section \ref{section:thenumericalstrategy} (in blue),
compared to a trace obtained with simple global rescaling of the steps (in red).
The sequences of signs in $\sigma_n=\pm\sqrt{\kappa\Delta_n}$ (that is, of left/right turns) 
are the same.}
\label{figure:chains}
\end{figure}

An example of a chain obtained by this method is presented in Figure \ref{figure:chains},
where it is compared with a chain obtained by non-random rescaling as in (\ref{eq:naivescaling}).

\section{\label{section:results}Asphericity and corrections to scaling}

Given a chain $\left\{\gamma_n\right\}$ --- both for a walk on the lattice and for a discrete SLE trace ---
one can define its \emph{gyration tensor}, which encodes useful information about the shape of the walk.
Since the process we have defined grows a chain towards infinity, it can not be compared to a whole
$N$-step walk on the lattice, because the latter displays finite-chain corrections close to its tip.
For this reason, we define the \emph{internal} gyration tensor, by following the definition of gyration tensor that is used
in polymer science, but by taking into account only the first $M$ monomers:
\begin{equation}
G^{\alpha\beta}(M)=\frac{1}{2 M^2}\sum_{i,j=1}^{M} \Big( \gamma_i^\alpha - \gamma_j^\alpha\Big)\Big( \gamma_i^\beta - \gamma_j^\beta\Big) ,
\end{equation}
where the superscripts $\alpha$ and $\beta$ take values on the $x$ and $y$ coordinates of a lattice site
or on the real and imaginary parts of a complex number.
Let $q_1(M)$ and $q_2(M)$ be the two (real) eigenvalues of the (symmetric) gyration tensor.
These quantities are not universal, but some of their combinations are believed to be, 
such as the \emph{asphericity}:
\begin{equation}
\mathcal A(M)=\left<\left(\frac{q_1(M)-q_2(M)}{q_1(M)+q_2(M)}\right)^2\right> ,
\end{equation}
which is a measure of how spherical the object is, 
being $0$ for perfectly spherical objects (for which the two eigenvalues are equal) 
and $1$ when one of the eigenvalues is $0$ (as happens for objects lying on a line).

The (critical) limit we are interested in is when the number of steps $N$ goes to infinity,
because this is the limit where the SAW displays its universal behavior and where the cut-off
at length 1 introduced by the disc-shaped forbidden region becomes irrelevant for
the discrete SLE.
On the SAW side, moreover, we will want to let $M\to\infty$, so as to avoid corrections to scaling, 
but in such a way as to have $M/N\to 0$, since we want to be looking at a portion deep inside the walk.

\begin{figure}[t]
\centering
\includegraphics[scale=0.9]{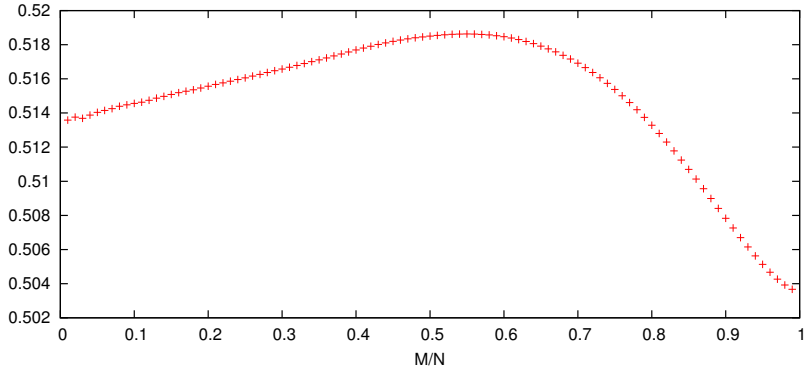}
\caption{(Color online) The internal asphericity of a SAW
as a function of the fraction of walk considered.
The error (not shown in figure) is $\approx 10^{-4}$ for each point
(about the size of the red crosses),
but this is to be taken \emph{cum grano salis},
since different points on the plot are obtained from the same
set of walks and are therefore not independent.
}
\label{figure:asphericity}
\end{figure}

We have simulated the self-avoiding walk on the square lattice
using the pivot algorithm \cite{MadrasSokal,Kennedy:pivot}
and we have measured the internal asphericity as a function of $M/N$
for $N=100\;000$ and $M=k\cdot 1000$ with $k=1,2,\ldots,99$.
The results are in Figure~\ref{figure:asphericity};
the curve plotted is expected to be universal, and to our knowledge has never appeared in the literature. 
In order to obtain the value at $M/N=0$ we perform a fit of the form $f(M/N)=\mathcal A_0^\mathrm{SAW}+ \alpha(M/N)^\theta$
on the first few values (namely $M/N\in(0,0.2]$) in the graph\footnote{
The results show a strong stability as the upper cut-off in $M/N$ is changed,  
up to around the middle of the chain, where they start to drift. }, 
obtaining
\begin{equation*}
\begin{split}
\mathcal A_0^{\mathrm{SAW}} &= 0.51343(5)\\
\big{[}\: \theta &= 0.92(5) \:\big{]} .
\end{split}
\end{equation*}

The same measure (now with $M=N$) is performed on ensembles of $20\;000$
discrete SLE traces ($\kappa=\frac{8}{3}$) of lengths $N=100,200, 400, 1000, 2000, 5000$,
(and $5000$ traces of length $10\;000$)
produced with the algorithm described in the previous sections.
The results are in Table~\ref{table:asphericity}\footnote{
We have also run simulations (for $N=100,200,400$) using a numerical solution
to (\ref{eq:transcendental}), for the sake of testing the approximation in (\ref{eq:Deltaj}).
The results agree with those in Table~\ref{table:asphericity}
[$0.4956(17)$, $0.5030(17)$, $0.5069(17)$ for $N=100,200,400$ respectively].}.
The simulation for $N=400$ took approximately one hour on a 2GHz Intel Duo processor.
\begin{table}
\caption{The asphericity of discrete SLE traces for $\kappa=\frac{8}{3}$,
computed for several lengths $N$.}
\label{table:asphericity}
\centering
\begin{tabular}{ll}
\hline\noalign{\smallskip}
$N$ & $\mathcal A^{\mathrm{SLE}}$\\
\noalign{\smallskip}\hline\noalign{\smallskip}
100 & 0.4957(17)\\
200 & 0.5019(17)\\
400 & 0.5071(17)\\
1000 & 0.5097(18)\\
2000 & 0.5115(17)\\
5000 & 0.5122(18)\\
10000 & 0.5124(37)\\
\noalign{\smallskip}\hline
\end{tabular}
\end{table}

In general, for an $N$-step chain one expects the following behavior
for the expectation value of a global observable $\mathcal O$:
\begin{equation}
\label{eq:correctionstoscaling}
\begin{split}
\frac{\left<\mathcal{O}\right>_N}{N^{p_{\mathcal O}}} = a &+ \frac{a_1}{N} + \frac{a_2}{N^2} + \cdots \\
&+ \frac{b_0}{N^{\Delta_1}} + \frac{b_1}{N^{\Delta_1+1}} + \frac{b_2}{N^{\Delta_1+2}} + \cdots \\
&+ \frac{c_0}{N^{\Delta_2}} + \frac{c_1}{N^{\Delta_2+1}} + \frac{c_2}{N^{\Delta_2+2}} + \cdots \\
&+ \cdots
\end{split}
\end{equation}
where the leading behavior (given by the exponent $p_\mathcal{O}$) is corrected
by analytical (with integer exponents) and confluent corrections.
The exponents\footnote{
The symbols $\Delta_1$, $\Delta_2$, $\ldots$ are those conventionally
used for these quantities: they do not have anything to do with the $\Delta_j$'s
used in previous sections.}
($\Delta_1<\Delta_2<\cdots$) are universal.
The asphericity is expected to be a constant in the large-$N$ limit, so that $p_{\mathcal A}=0$.
The scaling form (\ref{eq:correctionstoscaling}) then suggests
a fit of the form $f(N)=\mathcal A^{\mathrm{SLE}}+b_0/N^{\Delta_1}$, which yields
\begin{equation*}
\begin{split}
\mathcal A^{\mathrm{SLE}} &= 0.51351(54)\\
\Delta_1 &= 0.694(60) .
\end{split}
\end{equation*}
The asphericity is in perfect agreement with the one obtained for the SAW.
Our result for $\Delta_1$ agrees with the theoretical value ($\Delta_1=\frac{11}{16}=0.6875$)
obtained by conformal-invariance methods for polymers in good solutions \cite{Saleur}.
The evaluation of $\Delta_1$ for SAWs has been the subject of debate in the past decades.
In fact, the rich structure of corrections in (\ref{eq:correctionstoscaling}) makes it difficult to extract
precise values from numerical data.
There is now strong evidence for the absence of a leading term with exponent
$\Delta_1=\frac{11}{16}$ on the square lattice, 
the first non-null confluent contribution having exponent $\frac{3}{2}$ \cite{Caracciolo:corrections}.
Notice that this does not configure a violation of universality, since the amplitudes of the corrections
are model-dependent.
By including the first analytical correction (at next-to-leading order)
and fixing $\Delta_1$ to its theoretical value, we find
\begin{equation*}
\begin{split}
\mathcal A^{\mathrm{SLE}} &= 0.51356(44)\\
a_1 &= 0.003\pm0.362 ,
\end{split}
\end{equation*}
where the amplitude $a_1$ of the term $a_1/N$ is compatible with $0$.

\section{Conclusions and outlook}

We have studied a stochastic process in the complex plane, based on discrete Stochastic-Loewner evolution, which gives rise to chains $\left\{\gamma_n\right\}$ with approximately constant steps $|\gamma_n - \gamma_{n-1}|$.
The purpose was to build an algorithm for exactly sampling self-avoiding paths, by correctly reproducing the parametrization induced by the scaling limit of lattice models, namely self-avoiding walks.
The method is based on iterative composition of conformal maps, where each map acts by building a radial slit out of the unit circle, which will eventually become one of the steps of the discrete path.
Each step has to be rescaled according to the Jacobian of the map that evolves it.
This program encounters some technical hindrances --- essentially due to the fact that rescaling a step actually changes the Jacobian.
We showed that an alternative approach is possible, by keeping track of the second derivative of the map.
It turns out that the Hessian can be effectively computed in the framework of iterated conformal maps, since it can be expressed as a simple function of quantities already computed by the algorithm.

By exploiting the powerful correspondence existing between SLE and SAWs, the algorithm presented here produces completely independent samples of self-avoiding paths from the origin to infinity in the plane, whose parametrization is the desired one --- that corresponding to SAWs ---, and does so in an affordable way, with a complexity $O(N^2)$ for $N$-step chains.
This allows us to study parametrization-dependent observables of the SAW such as the internal asphericity and the leading correction-to-scaling exponent, whose determination is considered a challenging problem in the numerical study of polymers.
The results we obtain are very accurate.

The analysis has been carried out in the whole-plane radial geometry, but very little should be changed in order to adapt it to the half-plane chordal case, or to other restricted geometries of interest in polymer science.

Interesting questions remain open.
It is still not clear whether there exists a way of reproducing finite-chain effects by the foregoing techniques.
It would be useful for instance to produce the correct distribution of the \emph{end-point} of a SAW, which is fixed to infinity in the present study.
On the other hand, a great advance would be to translate this method to other classes of critical polymers, such as the $\theta$ point
where the collapsing transition poses even more difficult problems to Monte Carlo methods due to the attractive interactions and the consequently entangled shapes.

\section*{Acknowledgements}

The author wishes to thank Sergio Caracciolo and Andrea Pelissetto for suggestions and encouragement.

\appendix

\section{Computing the second derivative of $g_n$}

In this Appendix we give a formula for the modulus of the second derivative of $g_n$, 
which is needed in a crucial step of the algorithm.

The $n$-th composed map (\ref{eq:nthmap}) can be written as
\begin{equation}
\label{eq:laststepcomposition}
g_n(z)=g_{n-1} \circ R_n \circ \phi_n(z)
\end{equation}
in terms of the $(n-1)$-th map $g_{n-1}$, for $n>1$.
We recall that $R_n$ and $\phi_n$ are the $n$-th rotation and slit mapping respectively
and that $R_n(z)$ implicitly depends on the parameter $\delta_n$,
while $\phi_n$ depends on $\Delta_n$.
The second derivative of (\ref{eq:laststepcomposition}) reads
\begin{equation}
\begin{split}
g''_n=&\left( g''_{n-1}\circ R_n\circ\phi_n\right) \left(R'_n\circ\phi_n\right)^2 \left(\phi'_n\right)^2\\
&+\left(R'_n\circ\phi_n\right) \left(\phi''_n\right) \left(g'_{n-1}\circ R_n\circ\phi_n\right) .
\end{split}
\end{equation}
The latter expression simplifies when evaluating its modulus at $z=1$, because $\phi'_n(1)=0$ ---
a hallmark of the singularity of the Loewner map in 1 --- and
$\left| R'_n\circ\phi_n (z) \right|=1$.
One obtains
\begin{equation}
\label{eq:modulusofsecondderivative}
\left|g''_n(1)\right|=\left|\phi''_n(1)\right| \left|g'_{n-1}\circ R_n\circ\phi_n(1)\right| .
\end{equation}
Computing the first factor $\left|\phi''_n(1)\right|$ is just a matter of differentiating (\ref{eq:wholeplanemap})
two times with $\phi^{\mathbb D}_j$ given by  (\ref{eq:slitmap}), which yields
\begin{equation}
\label{eq:phiprime1}
\phi'_n(z)=\frac{\phi_n^{{\mathbb D}\:\prime}(1/z)}{z^2\left[\phi_n^{\mathbb D}(1/z)\right]^2}
\end{equation}
with
\begin{equation}
\label{eq:phiprime2}
\phi_n^{{\mathbb D}\:\prime}(z) =\frac{(z-1)\left[\re^{\Delta_n}(z+1)\left(\sqrt{(z+1)^2-4\re^{-\Delta_n}z} -z-1\right) + 2z\right]}{2z^2\sqrt{(z+1)^2-4\re^{-\Delta_n}z}} ,
\end{equation}
and
\begin{equation}
\label{eq:phisecond}
\left| \phi''_n(1)\right| = \frac{1}{2}\re^{\Delta_n}\left(2+\sqrt{1-\re^{-\Delta_n}}+\frac{1}{\sqrt{1-\re^{-\Delta_n}}} \right) .
\end{equation}
For the second factor $\left|g'_{n-1}\circ R_n\circ\phi_n(1)\right|$, instead, 
a closed recursion can be found by noting that, by differentiating (\ref{eq:laststepcomposition}), one has
\begin{equation}
\label{eq:recursion}
\begin{split}
\left|g'_{n-1}\right|&=\left|g'_{n-2}\circ R_{n-1}\circ\phi_{n-1}\right| \left|R'_{n-1}\circ\phi_{n-1}\right| \left|\phi'_{n-1}\right|\\
&=\left|g'_{n-2}\circ R_{n-1}\circ\phi_{n-1}\right| \left|\phi'_{n-1}\right| ,
\end{split}
\end{equation}
so that
\begin{equation}
\begin{split}
\left| g'_{n-1}\circ R_n\circ\phi_n(1)\right| =& \left| \phi'_{n-1}\circ R_n\circ\phi_n(1)\right|\cdot\\
&\left| g'_{n-2}\circ R_{n-1}\circ\phi_{n-1}\circ R_n\circ\phi_n(1)\right| .
\end{split}
\end{equation}
The seed of the recursion is given by the first slit grown
\begin{equation}
\label{eq:seedofrecursion}
g_1(z)=R_1\circ\phi_1(z) ,
\end{equation}
so that finally from (\ref{eq:modulusofsecondderivative}), (\ref{eq:recursion}) and the derivative of (\ref{eq:seedofrecursion}) one obtains
\begin{equation}
\begin{split}
\left|g''_n(1)\right| =& \left|\phi''_n(1)\right|\cdot\\
&\left|\phi'_{n-1}\circ R_n\circ\phi_n(1)\right|\cdot\\
&\left|\phi'_{n-2}\circ R_{n-1}\circ\phi_{n-1}\circ R_n\circ\phi_n(1)\right|\cdot\\
&\cdots\\
&\left|\phi'_1\circ R_2\circ\phi_2\circ R_3\circ\phi_3\circ \cdots \circ R_{n-1}\circ\phi_{n-1}\circ R_n\circ\phi_n(1)\right|
\end{split}
\end{equation}
or, in a more compact form,
\begin{equation}
\left|g''_n(1)\right|=\left|\phi''_n(1)\right| \prod_{j=0}^{n-2}\left|\phi'_{n-1-j}\left(\Gamma_j\right)\right|
\end{equation}
with $\Gamma_j$ defined as in (\ref{eq:Gamma}).


\begin{thebibliography}{99}
\bibitem{DesCloizeauxJannink} Des Cloizeaux, J., Jannink, G.: Polymers in solution: Their modelling and structure. Oxford University Press, New York (1990)
\bibitem{Schaefer} Sch\"afer, L.: Excluded volume effects in polymer solutions. Springer-Verlag, Berlin-New York (1999)
\bibitem{MadrasSlade} Madras, N., Slade, G.: The self-avoiding walk. Birkh\"auser, Boston (1993)
\bibitem{GuttmannConway} Guttmann, A.J., Conway, A.R.: Square lattice self-avoiding walks and polygons. {\it Ann. Comb.} {\bf 5} 319--45 (2001)
\bibitem{Sokal:montecarlomethods} Sokal, A.D.: Monte Carlo methods for the self-avoiding walk. {\it Nucl. Phys. B} {\bf 47} 172--9 (1996)
\bibitem{vanRensburg} Janse van Rensburg, E.J.: Monte Carlo methods for the self-avoiding walk. {\it J. Phys. A: Math. Theor.} {\bf 42} 323001
\bibitem{LSW:planarSAW} Lawler, G.F., Schramm, O., Werner, W.: On the scaling limit of planar self-avoiding walks. {\it Proc. Symp. Pure Math.} {\bf 72} vol~2 339--64 (2004)
\bibitem{Kennedy:lengthofanSLE} Kennedy, T.: The length of an SLE --- Monte Carlo Studies. {\it J. Stat. Phys.} {\bf 128} 1263--77 (2007)
\bibitem{Gherardi} Gherardi, M.: Whole-plane self-avoiding walks and radial Schramm-Loewner evolution: a numerical study. {\it J. Stat. Phys.} {\bf 136} 864--74 (2009)
\bibitem{Lawler:reparam} Lawler, G.F.: Dimension and natural parametrization for SLE curves. Preprint. arXiv:0712.3263v1 [math.PR]
\bibitem{LawlerSheffield:reparam} Lawler, G.F., Sheffield, S.: The natural parametrization for the Schramm-Loewner evolution. Preprint. arXiv:0906.3804v1 [math.PR]
\bibitem{Lawler:book} Lawler, G.F.: Conformally invariant processes in the plane. {\it Mathematical Surveys and Monographs} {\bf 114} American Mathematical Society (2005)
\bibitem{Werner:stFlour} Werner, W.: Random planar curves and Schramm-Loewner evolutions (Lecture notes from the 2002 Saint-Flour summer school). {\it L. N. Math.} {\bf 1840} 107--95 (2004)
\bibitem{Cardy} Cardy, J.: SLE for theoretical physicists. {\it Ann. Phys.} {\bf 318} 81-118 (2005)
\bibitem{KagerNienhuis} Kager, W., Nienhuis, B.: A guide to stochastic Loewner evolution and its applications. {\it J. Stat. Phys.} {\bf 115} 1149--229 (2004)
\bibitem{BauerBernard} Bauer, M., Bernard, D.: 2D growth processes: SLE and Loewner chains. {\it Phys. Rept.} {\bf 432} 115-221 (2006)
\bibitem{Bauer:DSLE} Bauer, R.O.: Discrete L\"owner evolution. {\it Ann. Fac. Sci. Toulouse VI} {\bf 12} 433--51 (2003)
\bibitem{Kennedy:reparam} Kennedy, T.: Monte Carlo comparisons of the self-avoiding walk and SLE as parametrized curves. Preprint. arXiv:math/0612609v2 [math.PR]
\bibitem{WittenSander:DLA} Witten, T.A., Sander, L.M.: Diffusion-limited aggregation: a kinetic critical phenomenon. {\it Phys. Rev. Lett.} {\bf 47} 1400--3 (1981)
\bibitem{HastingsLevitov:DLA} Hastings, M.B., Levitov, L.S.: Laplacian growth as one-dimensional turbulence. {\it Physica D} {\bf 116} 244--52 (1998)
\bibitem{Schramm} Schramm, O.: Scaling limits of loop-erased random walks and uniform spanning trees. {\it Israel J. Math.} {\bf 118} 221--88 (2000)
\bibitem{Hastings:multifractal} Hastings, M.B.: Exact multifractal spectra for arbitrary Laplacian random walks. {\it Phys. Rev. Lett.} {\bf 88} 055506 (2002)
\bibitem{MadrasSokal} Madras, N., Sokal, A.D.: The pivot algorithm: a highly efficient Monte Carlo method for the self-avoiding walk. {\it J. Stat. Phys.} {\bf 50} 109--86 (1988)
\bibitem{Kennedy:pivot} Kennedy, T.: A Faster implementation of the pivot algorithm for self-avoiding walks. {\it J. Stat. Phys.} {\bf 106} 407--29 (2002)
\bibitem{Saleur} Saleur, H.: Conformal invariance for polymers and percolation. {\it J. Phys. A: Math. Gen.} {\bf 20} (1987)
\bibitem{Caracciolo:corrections} Caracciolo, S., Guttmann, A.J., Jensen, I., Pelissetto, A., Rogers, A.N., Sokal, A.D.: Correction-to-scaling exponents for two-dimensional self-avoiding walks. {\it J. Stat. Phys.} {\bf 120} (2005)

\end{thebibliography}
\end{document}